\documentclass{svproc}
\usepackage{url}
\usepackage{graphicx}
\usepackage{amsmath}

\begin{document}
\mainmatter              
\title{Exploring Complexity Changes in Diseased ECG Signals for Enhanced Classification}
\titlerunning{Complexity Changes in ECG Signals}  
%
\author{Camilo Quiceno Quintero \and
Sandip Varkey George\inst{1}}
\authorrunning{Quintero \& George} 
%
%
\institute{Department of Physics, University of Aberdeen, Aberdeen AB24 3UE, United Kingdom\\
\email{sandip.george@abdn.ac.uk}}

\maketitle              

\begin{abstract}
The complex dynamics of the heart are reflected in its electrical activity, captured through electrocardiograms (ECGs). In this study we use nonlinear time series analysis to understand how ECG complexity varies with cardiac pathology. Using the large PTB-XL dataset, we extracted nonlinear measures from lead II ECGs, and cross-channel metrics (leads II, V2, AVL) using Spearman correlations and mutual information. Significant differences between diseased and healthy individuals were found in almost all measures between healthy and diseased classes, and between 5 diagnostic superclasses ($p<.001$). Moreover, incorporating these complexity quantifiers into machine learning models substantially improved classification accuracy measured using area under the ROC curve (AUC) from 0.86 (baseline) to 0.87 (nonlinear measures) and 0.90 (including cross-time series metrics).
\end{abstract}
\section{Introduction}

Cardiovascular diseases (CVDs) are the leading cause of death worldwide, accounting for almost 20 million deaths in 2019\cite{roth2020global}. This burden is disproportionately higher in low- and middle-income countries, underscoring the need for cost-effective tools for early detection and intervention for CVDs.  The electrocardiogram (ECG) is a powerful and cost effective diagnostic tool that can provide critical insights into heart rate and rhythm. The ECG measures the electrical activity of the heart which produces a characteristic waveform called the PQRS complex that repeats almost periodically. The exact form and timing of this PQRS complex varies between beats, resulting in a nonlinear and complex signal. While the overall shape of the PQRS complex is important to diagnose pathology, deviations in the complexity of the waveform from a baseline level is also associated with various conditions \cite{shekatkar2017detecting,gupta2022robust,george2023predicting}. 

ECG analysis typically involves standard measures such as power spectral features and inter-peak timing. However, quantifiers that measure the nonlinear dynamics of the signal such as entropy measures, fractal dimensions and recurrence quantification analysis (RQA) are also relevant to cardiac pathology \cite{costa2003multiscale,mathunjwa2021ecg,behera2024non,smigiel2021ecg}.  Beyond these single time series measures, the correlation and information overlap between the 12 ECG channels measures the the variations in the electrical activity over the body, and consequently gives insight into cardiac health. Commonly used cross-time series measurements include correlation coefficients, mutual information, transfer entropy and cross recurrence measures.

In the recent past there has been a steady increase in the available computational power as well as the availability of well annotated large time series datasets. This has lead to a surge in the popularity of machine learning algorithms for analysis. Moreover, the increased computational power has also made it possible to quantify complexity metrics for large time series datasets in a relatively short time. 

In this study we ran our analysis on the ECG data available from the PTB-XL data\cite{wagner2020ptb}. We initially calculated nonlinear measures from the lead II time series, namely the Higuchi fractal dimension (HD), Approximate Entropy (ApEn), Multiscale entropy (MSE) and recurrence plot (RP) measures, and cross time series measures between leads II, AVL and V2 time series, namely the Spearman correlation coefficient and mutual information. Differences between these for diseased and healthy individuals, as well as between 5 diagnostic super classes were quantified. Machine learning models—including logistic regression (LR), support vector machine (SVM), random forest (RF), XGBoost and artificial neural networks (ANN)-were then used to classify these data. Initially, we used a baseline model without complexity features. We then added single time series and cross-channel complexity metrics to improve classification accuracy. Moreover, we also conducted a five class classification into five diagnostic super classes.

\section{Methods}

\subsection{Data}
The PTB-XL ECG database was used for analysis in this study. The dataset consisted of $21,837$ clinical $12$-lead ECGs from $18,885$ patients. The recordings lasted 10 seconds each, and were sampled at 100 Hz with a 16-bit resolution, leading to time series of 1000 points each. Each ECG record in the dataset was annotated originally by upto two cardiologists with mutiple ECG statements. Based on these statements, the records were categorized into five diagnostic superclasses, namely Normal (NORM), Conduction Disturbance (CD), Myocardial Infarction (MI), Hypertrophy (HYP) and ST/T change (STTC) \cite{wagner2020ptb}. In this study, we further classify them into healthy and diseased binary classes, based on whether the ECG was categorized as NORM or into one of the diseased classes. Further details of the dataset may be found in \cite{wagner2020ptb}.

\subsection{Pre-processing}
\begin{figure}
    \centering
    \includegraphics[width=0.75\linewidth]{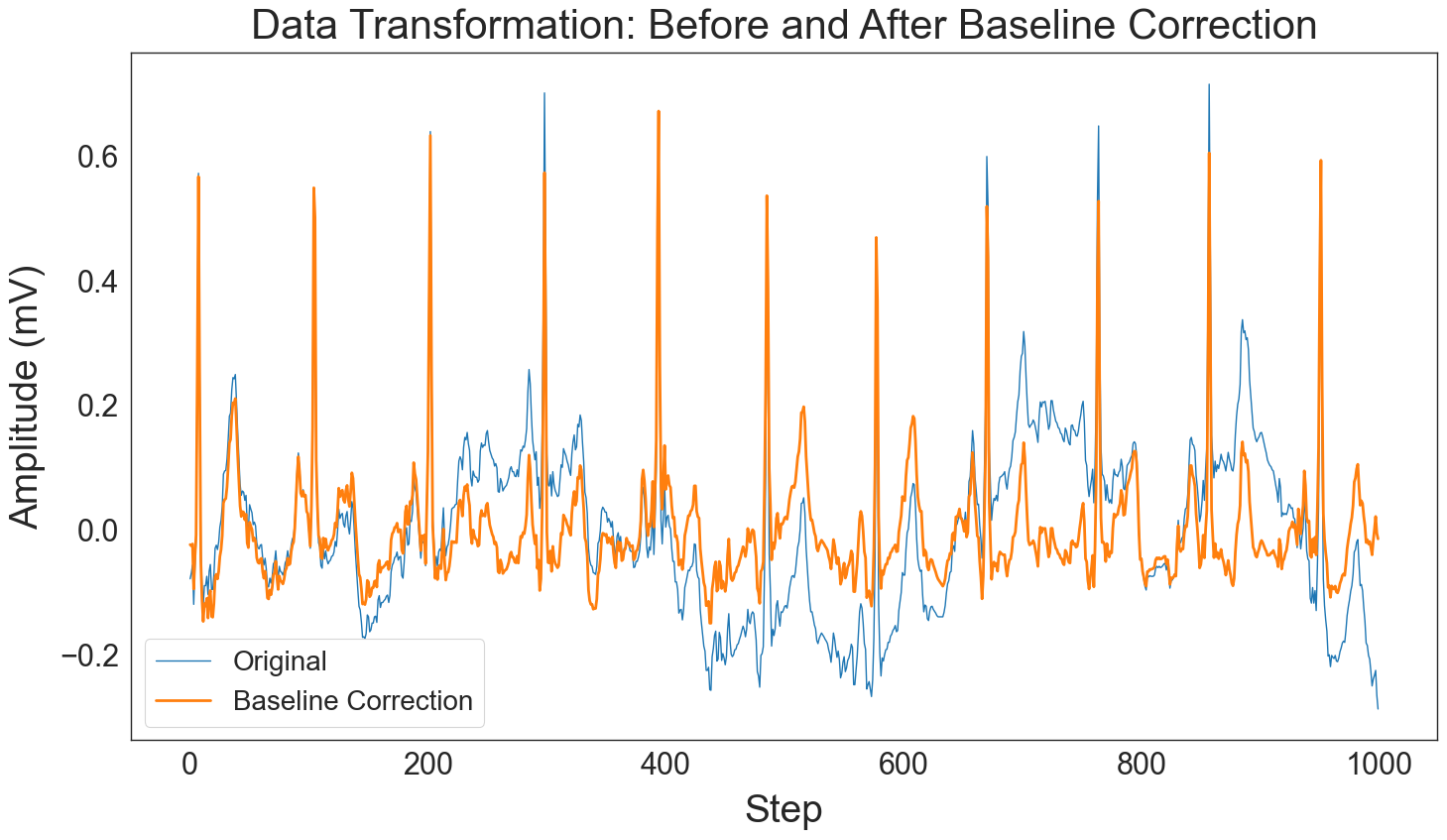}
     
    \caption{Figure illustrating the baseline correction on ECG data using a polynomial fit.}
    \label{fig:baseline_peakdetect}
\end{figure}

Many of the ECG time series exhibited a baseline drift, which was removed using a polynomial detrending method. A polynomial of order 20 was chosen based on \cite{shekatkar2017detecting} to balance the removal of drift without overfitting to signal fluctuations. This was subtracted from the original time series to leave a detrended residual behind. The original and detrended time series for a sample case is shown in Figure \ref{fig:baseline_peakdetect}. Since the resultant detrended ECG time series all exhibited differences in their amplitude ranges, they were standardized with a mean of $0$ and unit standard deviation.

\subsection{Single and cross time series metrics}
The pre-processed time series were then analysed using a number of single and cross channel metrics. Notably, we studied complexity metrics from the lead II ECG and the cross channel metrics between lead II, lead augmented vector left (AVL)  and lead V2 \cite{francis2016ecg}. 

\subsubsection{Complexity Analysis}
A range of complexity measures were computed for the lead II data of the ECG time-series. Lead II ECG is measured between the right arm and left leg leads, and is known to be a good lead to detect abnormalities in heart rate and rhythm \cite{Meek415}. The complexity measures that were calculated included:
\begin{enumerate}
    \item \textbf{Higuchi Fractal Dimension (HD)}: Quantifies the fractal dimension of time-series data, reflecting the self similarity at different scales in the signal and its  dimensional complexity\cite{higuchi1988approach,higuchi1990relationship}.
    \item \textbf{Entropy Measures}: Entropy measures broadly quantify the uncertainty or randomness in a time series. While there are a number of entropy measures that are popular in literature, we used the following to assess the disorder in the data.
    \begin{itemize}
        
       \item \textbf{Approximate Entropy (ApEn)}: Measures the regularity of fluctuations in a time-series by comparing repeating patterns in neighboring points. A lower for ApEn is indicative of higher predictability\cite{pincus1991regularity}.
        
        \item \textbf{Permutation Entropy (PermEn)}: Determines the complexity in a time series by evaluating the relative ordering of values within overlapping subsequences of dimension $m$. The entropy of the frequency distribution of these ordinal patterns is computed.
        
        \item \textbf{Lempel–Ziv Complexity (LZC)}: Converts the signal into a binary sequence and quantifies the range of unique patterns encountered in this sequence \cite{aboy2006interpretation}.
        
        \item \textbf{Multiscale Entropy (MSE)}: It extends the concept of entropy to measure the predictability of fluctuations at different temporal scales.

    \end{itemize}
    \item \textbf{Recurrence Plot-Based Measures}: The recurrence plot (RP) is constructed by first embedding the time series in a state space of dimension $m$ (chosen to be 3 in this study), through the method of delay embedding \cite{ambika2020methods}. Every point that recurs in the neighborhood of another in this embedded phase space is marked as a dark point in the RP, while every other is marked as white, resulting in a $n \times n$ plot \cite{marwan2007recurrence}. Sequences of neighboring points, diagonal and vertical dark lines are quantified using Determinism (DET), Laminarity (LAM), DET LAM ratio (DbyL),Entropy of Diagonal lines ($D_{ent}$), Entropy of Vertical lines ($V_{ent}$) and Trapping Time (TT).
\end{enumerate}
    
These selection of features were chosen to capture complementary aspects of the complexity of ECG signals. These were calculated using standard packages in Python v 3.11.3, namely the HFDA package for HD, the NeuroKit2 package for entropy measures and pyunicorn for RP measures \cite{makowski2021neurokit2,donges2015unified}.

\subsubsection{Cross time series measures}
The cross time series metrics were measures between leads II, aVL and V2. These leads were chosen, since they were shown to have the least redundancy among all the measured leads \cite{ramirezquantifying}. These metrics quantify inter-lead dependencies, providing complementary information to single-channel analyses. Two quantifiers were chosen as cross-time series metrics, namely the Spearman correlation coefficient ($\rho_S$) and mutual information (MI). When computed between the three leads, this results in a total of 6 quantifiers. $\rho_S$ quantifies the correlation between the leads. Unlike the Pearson correlation, the Spearman correlation also captures nonlinear relationships between variables as long as they are monotonic \cite{press2007numerical}. The MI captures the dependence between two time series based on the amount of information overlap between them\cite{dionisio2004mutual}.  

\subsection{Statistical testing}
The differences of the measures between the healthy and diseased classes, as well as the five disease superclasses were quantified using non-parametric tests, namely the Mann-Whitney u-test and the Kruskal Wallis test respectively. Non parametric tests were chosen, since they do not make assumptions on normality of the distributions of the tested populations. The Mann Whitney u-test tests for the null hypothesis that two samples come from the same distribution. The Kruskal Wallis test expands this hypothesis to test whether two or more independent samples arise from the same distribution \cite{gibbons2014nonparametric}. All statistical tests were tested for significance at an $\alpha$ level of $0.001$, and conducted using the scipy package in python \cite{virtanen2020scipy}.

\subsection{Machine learning classification}
The measures quantified above were used as input features to various machine learning algorithms to classify the ECG into disease classes. Two types of classification were conducted in this study, namely a two class classification into healthy and diseased classes, and a five class classification into five superclasses. The Logistic Regression (LR), Random Forest (RF), XGBoost (XGB) and Artificial Neural Network (ANN) were used as machine learning models for the binary classification, whereas only the ANN was used for 5 class classification.
The performance of the classifiers were quantified using the accuracy, the Matthew's correlation coefficient (MCC), and the area under the receiver operating characteristics curve (AUC). For the five class classification, in addition to the average MCC and accuracies, the one-vs-one and one-vs-rest AUC scores at a macro level and weighted by prevalence were also used. All the machine learning models were constructed using the scikitlearn package in python \cite{pedregosa2011scikit}. Codes for this study are available at \url{https://github.com/sgeorge91/PTB-XL-Complexity}.

\section{Results}
In the following subsections we present the results of our analyses.

\subsection{Statistical differences}
We initially checked for statistical differences between the healthy and diseased classes using Mann-Whitney u-tests. Every single channel quantifier showed significant differences between the two, except the MSE ($p<.001$), and every cross channel quantifier except $MI-V2-AVL$ showed significant differences ($p<.001$). The results of the u-tests, namely the value of the z-statistic and the p-value are shown in Table \ref{tab:mannwhitney_kruskalwallis}. The distributions with significant differences in the u-tests are marked with an asterisk. 

\begin{table}
    \centering
    \begin{tabular}{|c|c|c|c|c|}
    \hline
        Metric & z statistic & p-value & H statistic & p-value\\\hline
         HD & -16.68 & $<.001$ & 166.88 & $<.001$\\
         ApEn & -30.27 & $<.001$& 568.043 & $<.001$\\
         PermEn & -38.09 & $<.001$& 493.26 & $<.001$\\
         MSE & -1.22 & 0.220& 561.84 & $<.001$\\
         LZC& -52.06 &$<.001$& 1258.72 &$<.001$ \\
         DET& 11.70 & $<.001$& 186.26 & $<.001$\\
         LAM& 4.77 & $<.001$& 94.16 & $<.001$\\
         $\frac{DET}{LAM}$& -17.89 & $<.001$& 310.32 & $<.001$\\
         TT& 22.68 & $<.001$& 344.88 & $<.001$\\
         $D_{ent}$& 27.44 & $<.001$& 353.51 & $<.001$\\
         $V_{ent}$& 20.07 & $<.001$& 276.46 & $<.001$\\
         $\rho$-II-AVL& 42.20 & $<.001$& 646.68 & $<.001$\\
         $\rho$-II-V2& 47.20 & $<.001$& 1087.16 & $<.001$\\
         $\rho$-V2-AVL& 55.45 & $<.001$& 1457.20 & $<.001$\\
         MI-II-AVL& 4.69 & $<.001$& 699.05 & $<.001$\\
         MI-II-V2& 9.12 & $<.001$& 891.94 & $<.001$\\
         MI-V2-AVL& -1.14 & $0.255$& 485.83 & $<.001$\\\hline

    \end{tabular}
    \caption{z-statistics and p-values of the Mann-Whitney u-test for differences in nonlinear measures between the healthy and diseased classes, and H-statistics and p-values of the Kruskal-Wallis test for differences in nonlinear measures between diagnostic superclasses. All measures were significant at an $\alpha$ level of $.001$.}
    \label{tab:mannwhitney_kruskalwallis}
\end{table}

\begin{figure}[h]
    \centering
  
      \includegraphics[width=.75\linewidth]{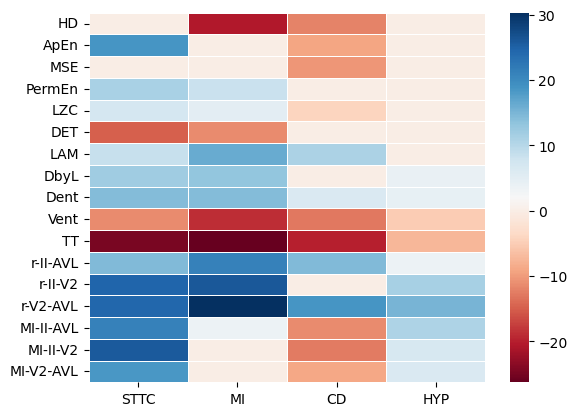} 
    \caption{Heatmap of the z-values from the pairwise Mann-Whitney u-test for various single and cross channel complexity measures, between each disease superclass the normal class. Z-values that are not significant at $\alpha=.001$ are set to $0$. Negative z values indicate the corresponding disease class has a higher median value of the measure as compared to the healthy class.}
    \label{fig:complex_heatmap}
\end{figure}

Subsequently, we check for differences between the 5 superclasses using the Kruskal Wallis test. All quantifiers showed significant differences between the classes ($p<.001$). The results for this are shown in Table \ref{tab:mannwhitney_kruskalwallis} To check for how these differ between the disease classes, we use pairwise u-tests. All tests were conducted between the 4 disease classes (STTC, MI, CD and HYP) and the NORM class. The z-values corresponding to the significant u tests are shown in Figure \ref{fig:complex_heatmap}. We see a difference in the magnitude and direction of the quantifiers between the disease classes, with STTC showing a lower value for most complexity quantifiers as compared to the diseased, whereas the CD class shows a higher value for most as compared to the diseased. The LAM, DbyL, $D_{ent}$ and the $\rho$ values were consistently lower across disease classes as compared to the normal, where as HD, $V_{ent}$, TT and DET showed consistently higher values across disease classes as compared to the normal. The other quantifiers varied between disease classes or were mostly not significant.

         
\subsection{Machine Learning models}
The above measures were used as features for a machine learning model for classifying the ECG data into binary classes and 5 superclasses. Initially, a number of standard features were generated from the time series to serve as a baseline model. These included the mean, median and standard deviation of the amplitude, the mean, median and standard deiviation of the distribution of RR peak intervals, the frequencies corresponding to the highest two powers in the power spectrum ($f_1,f_2$), the power of these frequencies ($P(f_1), P(f_2)$) and the ratio of powers of $f_1$ and $2f_1$ ($P(f_1)/P(2f_1)$).

\subsubsection{Two class classification}
\begin{figure}

      \includegraphics[width=.5\linewidth]{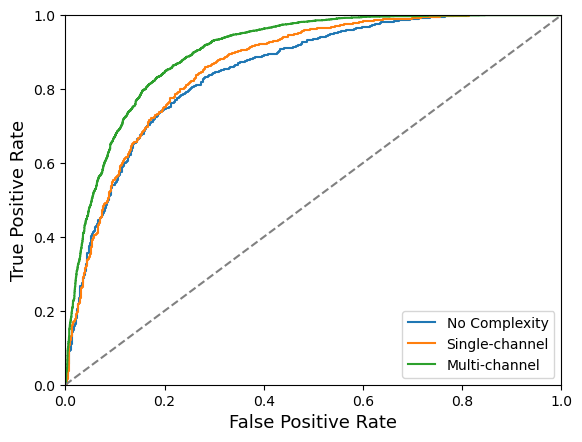} 
       \includegraphics[width=.5\linewidth]{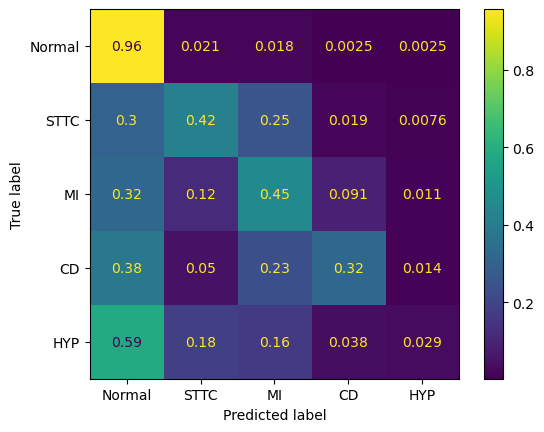} 
    \caption{Performance of the ANN classifier for 2 and 5 class classification. Left shows the ROC curve of the ANN classifier (a) in the absence of complexity metrics (blue),  (b) adding single channel complexity metrics (orange) and (c) adding cross channel complexity metrics (green). Right shows the confusion matrix for the 5-class classification ANN model.}
    \label{fig:roc_con}
\end{figure}

For the base models we used a feature set without any complexity or cross channel measures. These models showed fairly highest classification metrics, with even the lowest performing LR model showing an accuracy of 0.76. Subsequently we added single channel complexity measures. All models showed a marginal improvement in classification accuracy with these measures. Adding cross channel measures substantially improved classification metrics, with accuracy increasing by about $0.03$ in comparison to the single channel models. Finally, in an attempt to improve classification, we also added the age, sex and weights of the individuals, which resulted in a marginal improvement in all metrics the case of the ANN, but did not do so in most other models. The results of these models are shown in Table \ref{tab:binclass}. The ANN model performed best, with the baseline model achieving an accuracy of 0.78, AUC of 0.86, and a Matthew’s correlation coefficient (MCC) of 0.55. Adding complexity measures improved these values to 0.79, 0.87, and 0.57, respectively. Adding cross-channel measures further boosted performance, with accuracy reaching 0.82, AUC 0.90, and MCC 0.64, and adding meta data measures improved these to 0.84, 0.91 and 0.67. The ROC curves for the best performing ANN models are shown in Figure \ref{fig:roc_con}.
\begin{table}[t]
    \centering
    \begin{tabular}{c|c|c|c}
    \hline
        Model & Accuracy & MCC & AUC\\\hline
         \multicolumn{4}{c}{Standard Measures}\\\hline
        LR & 0.76  & 0.51 & 0.83\\
        RF & 0.77 & 0.54 & 0.84\\
        XGB &0.77  & 0.53 & 0.84\\
        ANN & 0.78 & 0.55 & 0.86\\\hline
         \multicolumn{4}{c}{Single channel Complexity}\\\hline
         LR& 0.77 & 0.53 & 0.84\\
         RF& 0.78 & 0.55 & 0.85\\
         XGB& 0.77 & 0.54 & 0.85\\
         ANN& 0.79 & 0.58 & 0.87\\\hline
         \multicolumn{4}{c}{Cross Channel Measures }\\\hline
         LR& 0.80 & 0.59 & 0.88\\
         RF& 0.81 & 0.62 & 0.89\\
         XGB& 0.81 & 0.62 & 0.88\\
         ANN& 0.82 & 0.65 & 0.90\\\hline
         \multicolumn{4}{c}{Metadata Measures }\\\hline
         LR& 0.81 & 0.60 & 0.86\\
         RF& 0.81 & 0.61 & 0.87\\
         XGB& 0.81 & 0.61 & 0.87\\
         ANN& 0.84 & 0.67 & 0.91\\\hline
         \hline
    \end{tabular}
    \caption{Performance metrics of the binary classification models, measured using the accuracy, MCC and AUC metrics. The ANN performed as the best performing model in every case. Addition of the single channel complexity metrics increased the performance marginally, while adding the cross channel metrics resulted in a higher improvement. Adding metadata measures, namely age, sex and weight, marginally improved the performance of some models, while it reduced the performance of others.}
    \label{tab:binclass}
\end{table}

\subsubsection{Five class classification}
Finally, we considered a five class classification using the best performing ANN model. The five class classification is considerably more challenging, with the model performance dropping drastically with an overall accuracy of $0.70$ and an MCC of $0.48$. A few different measures of AUC exist for multiclass classifications. For our model, the one-vs-one AUC scores were 0$.79$ at a macro level and $0.84$ when weighted by prevalence. Similar values for one-vs-rest AUC scores were $0.84$ at a macro level and $0.87$ when weighted by prevalence. The confusion matrix for the 5-class classification is shown in Figure \ref{fig:roc_con}. The lower values of metrics for the five-class classification demonstrates that this task is challenging. 
  
\section{Conclusions}
In this study we examined how the complexity of the ECG varies with cardiac pathology. Specifically, we considered how various measures of time series complexity computed from ECG data varies between diseased and healthy classes. We also study how these measures differ between different superclasses of disease. Broadly, we find that almost all complexity quantifiers are significantly different between the healthy and diseased classes (p-value$<.001$). This difference persists between the 5 disease superclasses. However the pairwise differences between them shows that the the complexity quantifiers differ in magnitude and direction of effect between different disease superclasses when compared with the normal class. A machine learning classification was also conducted on the data, using 4 different algorithms (LR, RF, XGB and ANN). The models were trained initially without any complexity measures, and then single and cross channel complexity features were added. The ANN model shows the best performance among all the machine learning algorithms for binary classification with the best performing model showing an average accuracy of $0.84$, and AUC of $0.91$. While single-channel complexity measures provided modest improvements to classification performance, integrating cross-channel measures yielded substantial gains, underscoring the importance of capturing inter-lead dynamics. Finally, a 5 class classification was performed using the ANN which gave an average accuracy across classes of $0.70$, and MCC of $0.48$.

This work is significant in multiple ways. It provides a comprehensive comparison of complexity and cross-channel measures in a large ECG dataset. It also establishes differences between different disease classes and shows how complexity of ECG varies between these classes. The significant differences observed across most complexity measures highlight their sensitivity to physiological changes in cardiac dynamics. For instance the striking differences in how complexity of ECG varies between STTC and CD classes highlights the need for tailored approaches to interpreting ECG patterns in various pathological contexts. We also show that adding complexity and cross channel features can enhance classification in simple machine learning models. Overall, our study provides a potential roadmap for future researchers to use complexity based features in their methods by showing the performance of different complexity metrics for different disease types, and also demonstrating how these metrics can enhance classification in machine learning methods.
\subsection*{Acknowledgements}
This work is funded by a Royal Society Research Grant awarded to SVG ($RG\backslash R1\backslash 241044$). This preprint has not undergone peer review or any post-submission improvements or corrections.
\bibliographystyle{spphys} 
\bibliography{nodyconbibfile}
%
%







\end{document}